\documentclass[aps,prl,groupedaddress,showpacs,twocolumn,floatfix]{revtex4}
\usepackage{amsmath}
\usepackage{amssymb}
\usepackage{graphicx}
\usepackage{epsfig}

\begin{document}

\title{Critical fields for vortex expulsion from narrow superconducting strips}
\author{P. S\'anchez-Lotero and J. J. Palacios}
\affiliation{\mbox{Departamento de F\'{\i}sica Aplicada, Universidad de Alicante,
\\ San Vicente del Raspeig, Alicante 03690, Spain}
}

\begin{abstract}
We calculate the critical magnetic fields for vortex expulsion for an infinitely long superconducting strip, using the Ginzburg-Landau formalism. Two critical fields can be defined associated with the disappearance of either the energetic stability or metastability of vortices in the center of the strip for decreasing magnetic fields. We compare the theoretical predictions for the critical fields in the London formalism  with ours and with recently published experimental results. As expected, for narrow strips our results reproduce better the experimental findings.
\end{abstract}

\maketitle
\narrowtext

\section{I. Introduction}
The Meissner effect\cite{Meissner}, i.e., the complete expulsion of an external magnetic field when a superconductor is cooled below the critical temperature, $T_c$, is a fundamental property of the  superconductivity.  The ratio between the superconducting characteristic lengths, $\lambda$ and $\xi$ (penetration and coherence length, respectively), allows for a classification of the superconductors as type-I or type-II. The type-I ones ($\kappa=\lambda / \xi < 1/\sqrt{2}$) exhibit a complete  Meissner effect below a critical field. The existence of impurities, grain boundaries, or simply geometrical effects can, nevertheless, favor the penetration of the magnetic field inside a type-I sample. Instead, type-II superconductors allow intrinsically for the penetration of magnetic field filaments with quantized flux (vortices) which form a triangular lattice\cite{Abrikosov}. 

The behavior of a vortex in a superconducting film  was first described by Pearl\cite{Pearl} in the London limit. He found the penetration length for a thin film with thickness $d$ to be described by $\Lambda=\lambda^2/d$. Also using the London approach for an ideal semi-infinite superconductor Bean and Livingston\cite{Bean} first showed that a surface barrier is produced as as result of the competition between the attraction of the vortex towards the surface due to an image antivortex outside the superconductor and the repulsion of the vortex due to surface screening currents that circulate in opposite direction to the vortex currents. Geometrical barriers for thin flat superconductors were introduced trying to explain the magnetization of high critical temperature superconductors\cite{Zeldov}. In the London limit various other problems such as the behavior of the vortex lattice in a type-II superconductor strip with current\cite{Brandt}, the energetics of a vortex near the edge of a thin superconducting film at zero field\cite{Kogan}, and the study of the Bean-Livingston barrier\cite{Kuznetsov} have been considered.

Whether or not vortices can be found inside a type-II superconductor and the density of them depends on whether the superconductor is field cooled or zero field cooled before applying the magnetic field. The fact that the number of vortices is not a unique number as a function of temperature and field is related, particularly in mesoscopic superconductors, to the existence of surface barriers for vortex escape and entrance\cite{Geim:nature:98,Palacios:prl:00}. The use of superconducting strips is common in some technological applications. Therefore, the characterization of this type of geometry is necessary for the design of superconducting devices. The noise generated by the movement of vortices causes loss of coherence in qubits and also affects the sensitivity of devices as the SQUIDS\cite{Bartolome}. In a recent work, Stan {\it et al.}\cite{Stan} have measured the minimum field below which the vortices are completely expelled from bidimensional Nb superconducting strips. The strips are cooled down through the critical temperature in the presence of a magnetic field applied previously. When the magnetic field is greater than $B_m$, vortices can be observed by a Scanning Microscopy Hall probe. For lower magnetic fields, vortices are completely expelled. 

From a theoretical point of view it makes sense to define two critical fields which are associated with the
disappearance of both the energetic stability and metastability of vortices in the center of the strip for decreasing magnetic fields. While an upper critical field can be obtained simply in terms of ground-state considerations, a careful computation of the surface energy barrier in the determination of the lower critical field is mandatory. In this work we study the nature of these two fields in narrow strips with the help of Ginzburg-Landau theory. We first review previous work based on the London model in Sec. II. In Sec. III we review the basics of Ginzburg-Landau theory applied to superconducting strips. In Sec. IV we make use of the latter approach and compare the theoretical predictions for the critical fields obtained in the London formalism with ours and with recently published experimental results. As expected, for narrow strips our results reproduce better the experimental findings. The conclusions are presented in Section V.

\section{II. Critical fields in the London limit}
Following the Bean-Livingston model\cite{Bean}, the free energy for a vortex in a strip as a function of the
transversal position $x$ can be described by\cite{Clem,Stan}
\begin{equation}
G(x) = \frac{4 \pi}{\mu_0} \left\{ \frac{\Phi_0^2}{4 \pi^2 \Lambda} \ln  \left[ \frac{2 w}{\pi \xi} \sin \left( \frac{\pi x}{w} \right) \right] - \frac{\Phi_0 B}{2 \pi \Lambda} x (w-x)\right\} 
\label{london}
\end{equation}
where $\mu_0$ is the permittivity of vacuum, $\Phi_0$ is the flux quantum, $B$ is the applied field, and $w$ is the width of the strip. Similar expressions for the vortex energy in a strip have been derived in Refs. \onlinecite{Kuznetsov,Kogan}. The first term corresponds to the nucleation energy of the vortex and the second to the interaction with the screening currents. Equation \ref{london} diverges for $x=0$ and is not valid within a distance $\xi$ from the edges. This is not surprising since it is based on the London approximation which neglects the core of the vortex. From the previous equation two critical fields, $B_0$ and $B_s$, can be defined. For low magnetic fields the free energy $G(x)$ displays a maximum in the center of the strip. As the magnetic field increases the maximum disappears at a field $B_0$
\begin{equation}
B_0= \frac {\pi \Phi_0}{4 w^2}.
\end{equation}
where $G(x)$ presents a plateau around $x=0$. For $B>B_0$ a local minimum appears in $x=0$ and a vortex can be placed there in a metastable situation. At a magnetic field $B_s>B_0$
\begin{equation}
B_s= \frac{2 \Phi_0}{\pi w^2} \ln \left( \frac{2 w}{\pi \xi} \right)
\end{equation}
the minimum becomes an absolute minimum of energy. Above this field the free energy of the strip with a vortex is lower than the free energy without vortices. The fields $B_0$ and $B_s$ are  two possible critical fields for the expulsion of the vortices. Arguments have been put forward in favor of $B_0$ as the relevant field for the total vortex exclusion\cite{Clem,Maksimova} and in favor of $B_s$\cite{Likharev,Stan}.

Experimental details on the manufacture and the set-up to observe the vortices in strips as well as measurements of $B_m$ are described in Ref. \onlinecite{Stan}. The strips had a thickness of $d=$210 nm, a length of $L_y=$4 mm, and a width of $w=$1.6, 10 and 100 $\mu$m, with a critical temperature, $T_c$=8.848 K. In order to find the values of $B_m$ for every strip, Stan {\it et al.} count the number of vortices for different values of the magnetic field after cooling down the samples through $T_c$. Their results show a linear dependence between the number of vortices and the magnetic field for higher fields. For low values of the field the dependence of the number of vortices with the field deviates from linearity until the vortices are completely expelled from the strip. The value of $B_m$ is found by extrapolating the linear zone. The working temperature $T_f$ is mantained close to $T_c$ where the mobility of the vortices allows them to explore the lowest possible energy configurations without getting pinned by impurities. This temperature is $T_f=8.835 K$. Taking $\xi_0$=38.9 nm and $\kappa=$5.0, $\xi(T_f)=$320 nm and $H_{c2}(T_f)=$ 3.217 mT. Thus, the narrowest strip has an approximate width of 5$\xi$. For the strips with width 10 $\mu$m, the experimental results of $B_m$ and the theoretical prediction of $B_s$ based on Eq. \ref{london} are basically in agreement with each other. For the widest strip $\Lambda \ll w$, the theoretical $B_s$ must be corrected by a factor of three to account for the partial field screening. However, for the narrowest strip, the London theory badly fails. For example, the theory predicts $B_0>B_s$ when the strips are narrower than $5\xi$. For narrow strips with widths of the order of $\xi$ the finite size of the vortex core cannot be ignored which is likely to be the reason behind the failure of the theory and the disagreement with the experimental results. For this reason we propose to describe the narrow strips using Ginzburg-Landau theory.

\section{III. Ginzburg-Landau theory in the lowest Landau level approximation}
The superconducting density is found by minimizing the Ginzburg-Landau functional\cite{Ginzburg}
\begin{equation}
\Delta G=\int dV \left[\Psi^* \hat{\Pi} \Psi +\alpha |\Psi|^2 + \frac{\beta}{2} |\Psi|^4 +\frac{[\vec{b}-\vec{H}]^2}{8 \pi} \right],
\end{equation}
with $\Delta G=G_s-G_n$ where $G_s$ and $G_n$ are the Gibbs free energies for the superconductor and normal state,
respectively, and $\hat{\Pi}=\frac{1}{2 m^*}\left( -i \hbar \vec{\nabla} - \frac{e^*}{c} \vec{A} \right)^2$ is the kinetic operator.
\begin{figure}
\centering
\includegraphics[scale=0.8]{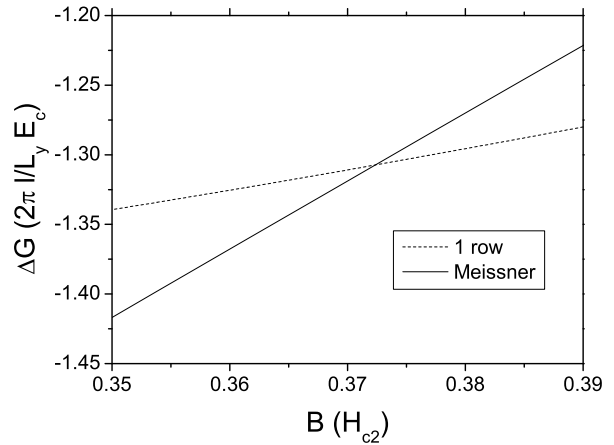}
\caption{Ginzburg-Landau functional for a superconducting strip of width 5 $\xi$. The transition of the ground state from Meissner (solid line) to 1 vortex row state (dashed line) occurs at 0.372 $H_{c2}$.
\label{graph4}}
\end{figure}
When the magnetic field is applied perpendicular to the superconducting thin film (situated on the $ xy$ plane), the effective penetration length $\Lambda$ can become greater than the width of the sample $w$, particularly close to $T_c$. In this case the screening of the magnetic field is weak and the field penetrates almost uniformly so we assume $\vec{b}=\vec{H}=B\hat{z}$ where $B$ is  spatially uniform. From now on we use the magnetic length $l=\sqrt{\frac{\hbar c}{e^* B}}$ and the cyclotron energy $E_c=\frac{\hbar\omega_c}{2}$ as length and energy units. The cyclotron frequency, is $\omega_c=e^* B/m^* c$.

We now expand the order parameter as a linear combination of eigenfunctions, $\phi_k$, of the kinetic operator with eigenenergies $\epsilon_k$. Choosing the Landau's gauge for the vector potential $\vec{A}=Bx\hat{y}$, one finds the following expression for the eigenfunctions:
\begin{equation}
\phi_k=Ae^{iky}\chi_k(x),
\end{equation}
where $k$ is the set of wavevectors in the direction $y$, $\chi_k(x)$ are functions without nodes, and $A$ is a normalization factor. These functions are calculated numerically and obey the usual boundary condition for a superconductor-vacuum system. In the lowest Landau level\cite{Landau} the free-energy functional becomes
\begin{eqnarray}
\Delta G &=&\sum_{k}^{N_c} |C_{k}|^2 \alpha_k + \frac{\beta}{2} \sum_{k_1,k_2,k_3,k_4}^{N_c} C_{k_1}^{*}C_{k_2}^{*}C_{k_3}C_{k_4} \times \nonumber \\
& &\int dx \chi_{k_1} \chi_{k_2} \chi_{k_3} \chi_{k_4} \delta_{k_3+k_4,k_1+k_2}
\end{eqnarray}
with $\Delta G$ expressed in units of $2\pi \frac{l}{L_y} E_c$ and $E_c=\frac{\hbar \omega_c}{2}$, $\beta$ in units of $E_c V$. $L_y$ is the length of the strip ($L_y \rightarrow \infty)$ and $N_c$ is the number of terms in the linear combination. $\alpha_k=\alpha+\epsilon_k$ is the condensation energy of the term $k$ in the expansion.
Details of the minimizing process can be found in Ref. \onlinecite{Palacios:prb:98}.
\begin{figure}
\centering
\includegraphics[scale=0.35]{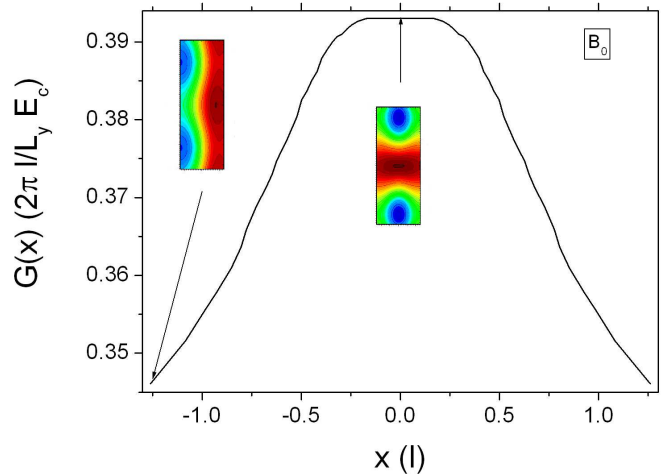}
\caption{One vortex row energy as a function of the vortex position across the strip where the energy of the Meissner state has been substracted. The magnetic field is $B_0=$0.265$H_{c2}$ and the width $5\xi$. The position is given in magnetic lengths. The insets show the superconducting density when the vortices are at the center and on the edge.
\label{B_0}}
\end{figure}

\section{IV. Critical fields in the Ginzburg-Landau formalism}
We determine $B_s$ by finding the magnetic field at which the Gibbs free energies of a system with $N_c=2$ (one vortex row) and $N_c =1$ (Meissner state) are equal. After minimizing the Ginzburg-Landau functional, we found for the narrowest strip a value of $B_s=$0.372$H_{c2}$ as shown in Fig. \ref{graph4}. For higher magnetic fields the penetration of another vortex row in the strip is not possible. For $N_c=2$ (one vortex row) the diference between wavevectors in the expansion, $\Delta k=k2-k1$, establishes the physical distance between vortices in the row, $\Delta y$,
\begin{equation}
\Delta y = \frac{2 \pi}{\Delta k}.
\end{equation}
For example the wavevectors at $B_s$ are $k=\pm$0.815$l^{-1}$ and therefore the distance between vortices is 3.85$l$. The difference between wavevectors changes with the magnetic field, therefore we have to search for the wavevectors that minimize the functional for every magnetic field.

By varying the values of the wavevectors, but keeping $\Delta k$ fixed, we can calculate the free energy of one vortex row with respect to the Meissner state, $G(x)$, as a function of the vortex row position across the strip. The variation of the wavevectors keeping their difference constant implies the numerical variation of coefficients and the minimization process turns to be completely numerical. In general for any configuration the coefficients of the linear combination are different between them and they are equal only when the vortex row is in the center of the strip. The displacement of a vortex row across the strip is associated with a current induced in the strip. In the Ginzburg-Landau formalism the existence of current in a superconducting strip with a vortex row increases the vortex linear density\cite{PhysicaC} and the distance between wavevectors in the expansion changes. Here we neglect this effect, which is small anyway.

\begin{figure}
\centering
\includegraphics[scale=0.35]{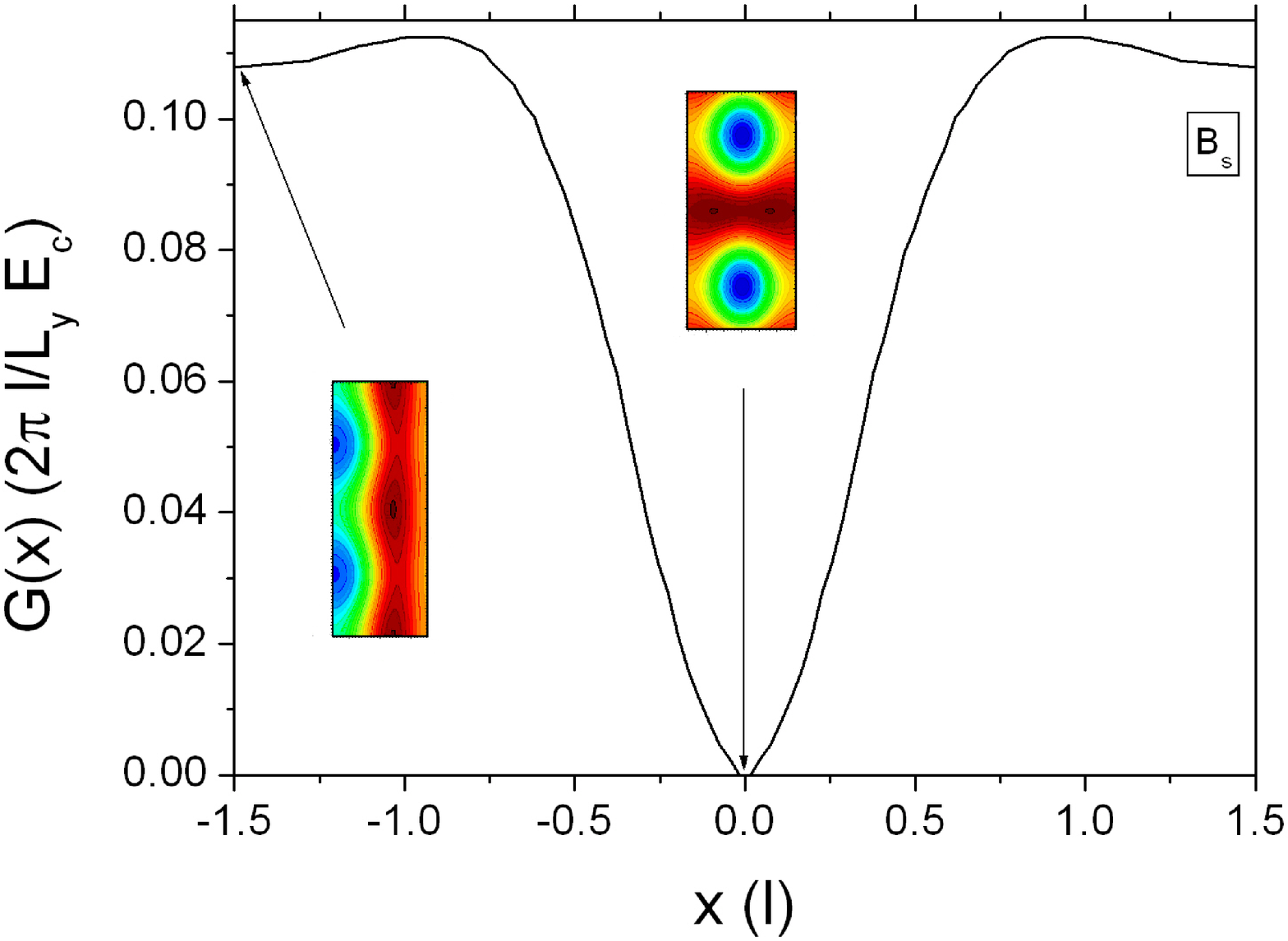}
\caption{One vortex row energy as a function of the vortex row position across the strip where the energy of the Meissner state has been substracted. The magnetic field is $B_s=$0.372$H_{c2}$ and the width $5\xi$. The position is given in magnetic lengths. The insets show the superconducting density when the vortices are at the center and on the edge.
\label{B_s}}
\end{figure}

We determine the critical field $B_0$ calculating $G(x)$ for various magnetic fields. $B_0$ is determined
by the appearance of  a plateau in the free energy in the center of the strip or the reversal of the sign of the
second derivative $d^2G(x)/dx^2\|_{x=0}$.  We find $B_0=$0.265$H_{c2}$ (see Fig. \ref{B_0}). At $B_0$ the distance between vortices is 4.269$l$ (see inset). As expected, this distance is greater than that at $B_s$ due to the increasing vortex density at higher fields. At $B=H_{c2}$ the distance between vortices is reduced to 1.872$l$, but have not disappeared yet due to the presence of the surfaces. For magnetic fields above $H_{c2}$ the vortices 
merge creating a channel without superconductivity in the center of the strip\cite{Palacios:prb:98}. We also compute $G(x)$ for the critical field $B_s$ (see Fig. \ref{B_s}). As expected a zero-energy mimimum is obtained in the center of the strip for which the superconducting density is shown in the inset.

The free energies curves at both fields $B_0$ and $B_s$ show a finite value when the vortices are placed on the edge. This behavior is different from that predicted by Eq. \ref{london} and can be attributed to a correct description of the order parameter in the Ginzburg-Landau calculation. The free energy for a vortex on the edge is superior to the one of the Meissner state in both cases. The position of the vortex is associated with the position of the center of its core although the core can be partially inside the strip. For instance, when the center of the core is exactly placed on the edge of the strip, the vortex has half of the vortex inside the strip and the other half outside (see insets in Figs. \ref{B_0} and \ref{B_s}). We note finally that the free energy as a function of the position of the vortex presents a maximum near the surface for $B_s$ similar to the one obtained in the London approximation\cite{Bean,Kuznetsov,Clem}. If a vortex nucleates at the surface still has to surpass the small barrier before it slides down towards the center of the strip.

\begin {figure}
\centering
\includegraphics[scale=0.85]{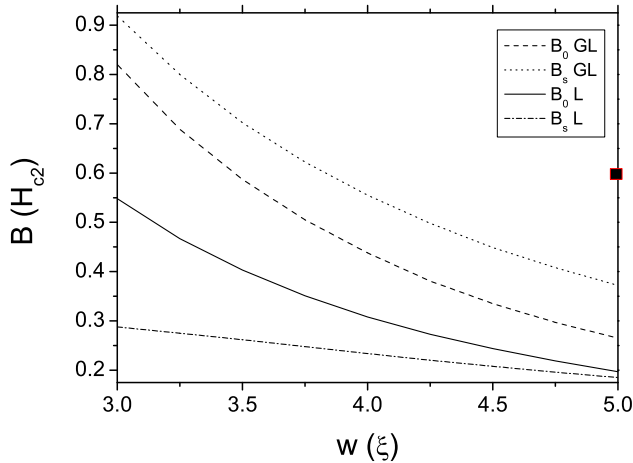}
\caption {Critical fields, $B_s$ and $B_0$, in the Ginzburg-Landau and London formalisms for different widths of a superconducting strip. The theoretical result of $B_s$ in the Ginzburg-Landau formalism is the closest value to the experimental value (solid square).
\label {graph1}}
\end {figure}

We now repeat the calculation of the critical fields $B_0$ and $B_s$ for widths between 3$\xi$ and $5\xi$. We find the expected behavior of the critical fields for all the widths, namely $B_s> B_0$, as shown in the Fig. \ref{graph1}. The closest result to the experimental data for the 5$\xi$ strip corresponds to the critical field $B_s$ using the formalism of Ginzburg-Landau. This result is compatible with the field cooled measurements of Stan et al. \cite{Stan}. Nevertheless, one should not forget that if the magnetic field decreases below $B_s$ the vortices would stay in the strip until the critical field $B_0$ is reached due to the saddle point energy barrier between the vortex solution and the Meissner solution. This barrier avoids the transition between the vortex states and the Meissner state despite of the former being energetically favorable. Thus, although the ground state for magnetic fields lower than $B_s$ has no vortices, the barrier would keep the vortices inside the strip until the critical field $B_0$ is reached. A combination of other effects not considered in this work such as thermal fluctuations and defects on the edge would decrease the escape barrier and would favor $B_s$ over $B_0$ also in decreasing fields.

\begin{figure}
\centering
\includegraphics[scale=0.84]{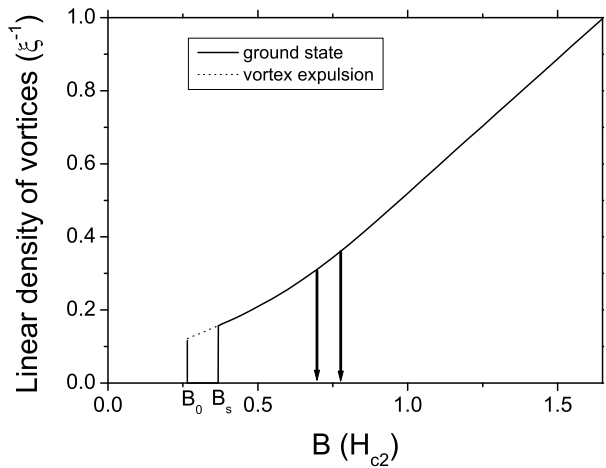}
\caption {Linear density of vortices for a 5$\xi$ superconducting strip. Solid and dashed lines correspond to the stable and metastable one vortex row state, respectively. The system can sustain vortices inside the strip for magnetic fields lower than $B_s$. Arrows delimit the field range used for the extrapolation of $B_m$ in Ref. \onlinecite{Stan}.
\label {hc1}}
\end {figure}

An almost linear behavior for the density of vortices as a function of the magnetic field can be seen in Fig. \ref{hc1} for large values of $B$. We also find an almost linear behavior for the density of vortices as function of the magnetic field for wider strips and large fields with the exception of small steps produced by the entrance of individual vortex rows (not shown). The ground state (solid line) has no vortices for magnetic fields lower than $B_s$ although, as stated in previous paragraph, the system could evolve in a metastable state (dashed line) until $B_0$ is reached. The arrows signal the region used by Stan et al. in their estimation of $B_m$\cite{Stan}. For lower fields we observe a deviation from linearity which also seems to be the case in the Stan {\it{et al.}} experiment, although the authors attribute this to imperfections or pinning.  We find a density of vortices greater than the experimental data, however the number of vortices expelled from the strip in the experiment is in agreement with the number the vortices expelled in our calculation in the same range of fields. The differences could be attributed to impurities or defects on the edges which are not accounted for in our simple geometry.

Finally we would like to note that the results for the critical fields present a similar behavior to the one predicted by London theory. According to London theory the expulsion field, $B_m$, can be approximated to $B_m \approx k/w^2$, where $k$ is a constant with a value close to $\Phi_0$. In our case the same expression applies, but with an exponent for the denominator of 2.216.

\section{V. Conclusions and discussion}
Two expulsion critical fields, $B_0$ and $B_s$, below which metastable and stable vortices, respectively,
cease to exist have been determined for infinitely long superconducting strips and widths between 3$\xi$ and 5$\xi$ using Ginzburg-Landau theory. The expected behavior for the values of the fields is obtained, always being
$B_0< B_s$ for this range of widths. The profile across the strip of the free energy of a vortex for the field $B_0$ presents a flat zone in the center of the strip whereas the one for $B_s$ has a zero-energy minimum for vortices at the center. The theoretical value of the two critical fields are closer to the experimental one than those obtained in the London formalism. Our value for $B_s$ for a strip with a width of 5$\xi$ is the closest value to the experimental one, as suggested by Stan et al.\cite{Stan} for wider strips. An overall agreement between our results for the number of vortices as a function of the field and the experiment is obtained, including the low field deviation from linearity at small values. If the observed deviation is not due to imperfections or defects, the experimental value for the critical field would be smaller and in closer agreement with our calculations. Finally, one should keep in mind that, in perfect samples free of imperfections, and ignoring temperature effects, the critical field for the expulsion of the vortices should correspond to $B_0$, at least in decreasing fields.

\section{Acknowledgements}
P. S-L. would like to thank to the Servicio de Relaciones Internacionales of the Universidad de Alicante, 
Caja de Ahorros del Mediterraneo and Instituto Alicantino de Cultura Juan Gil-Albert for its support. 
This work has also been funded by the Spanish MCYT under grant No. FIS2004-02356.

\end{document}